\documentclass[a4paper]{JHEP3}
\usepackage{epsfig,multicol,bbm}

\newcommand{\be}{\begin{equation}}
\newcommand{\ee}{\end{equation}}
\newcommand{\bea}{\begin{eqnarray}}
\newcommand{\eea}{\end{eqnarray}}
\newcommand{\mbb}{\mathbb}

\newcommand{\half}{\frac{1}{2}}
\newcommand{\mc}{\mathcal}
\newcommand{\gsim}{\gtrsim}

\title{Scanning the Landscape of  Flux Compactifications: Vacuum
  Structure and Soft
  Supersymmetry Breaking}
\author{Shehu S. AbdusSalam, Joseph P. Conlon, Fernando Quevedo, Kerim
  Suruliz\\ DAMTP, Centre for Mathematical Sciences,\\ Wilberforce
  Road, Cambridge, CB3 0WA, UK.\\
E-mails: \email{s.s.abdussalam@damtp.cam.ac.uk},
 \email{j.p.conlon@damtp.cam.ac.uk},
 \email{f.quevedo@damtp.cam.ac.uk}, \email{k.suruliz@damtp.cam.ac.uk}
}
\preprint{DAMTP-2007-79}

\abstract{We scan the landscape of flux compactifications for the
Calabi-Yau manifold $\mathbb{P}^4_{[1,1,1,6,9]}$ with two K\" ahler
moduli by varying the value of the flux
superpotential $W_0$ over a large range of values. We do not include uplift terms. 
We find a rich phase
structure of AdS and dS vacua. Starting with $W_0\sim 1$
we reproduce the exponentially large volume scenario, but as $W_0$ is
reduced new classes of minima appear. One of them corresponds to the
supersymmetric KKLT vacuum while the other is a new, deeper
non-supersymmetric minimum. We study
how the bare cosmological constant and the soft supersymmetry
breaking parameters for matter
on D7 branes depend on $W_0$, for these classes of
minima. We discuss potential applications of our results.}

\begin{document}

\section{Introduction}

Originally it was hoped that, once non-perturbative effects were understood, string theory would have
a unique vacuum. Finding this vacuum would allow all low-energy physical parameters to be directly computed.
However, as our understanding developed this view has come to be seen as naive. There instead seems to be a landscape
of many different vacua, each having different physical properties. Different vacua may have different gauge groups,
different particle representations, and different values for the cosmological constant. The vacuum that we inhabit may be just one
of a very large set of possibilities.

The construction that puts this problem in sharpest focus are flux compactifications, in which the RR and NS-NS
fields present in string theory take non-trivial profiles in the
vacuum
(for reviews see \cite{fluxcomp}). Fluxes take integral charges within the co-homology
lattice of the compactification manifold. As typical Calabi-Yaus may have $\mc{O}(100)$ cycles, the number of different profiles
the fluxes can be given is exceedingly large. Each choice of fluxes correspond to a different vacuum state. As the fluxes enter the
low-energy Lagrangian, for example through the dynamics of moduli fields, there exist very many different low energy Lagrangians.

The landscape of low-energy theories removes the hope of a unique determination of the vacuum state. However, it is 
fortunately not the case that the 
low-energy theories are entirely arbitrary. In IIB compactifications, the fluxes stabilise the dilaton and complex structure moduli
through the superpotential \cite{GVW, GKP}:
\be
W = \int G_3 \wedge \Omega.
\ee
However the direct effect of the fluxes on the dynamics of the K\"ahler moduli is relatively limited, and is mostly determined by the
vev of the above flux superpotential. Different vacua have been identified depending on different values of the flux superpotential $W_0$.
For example, when $\vert W_0 \vert \sim 1$ there exist non-supersymmetric vacua at exponentially large values of the volume
(the large volume models \cite{hepth0502058})
whereas for $W_0 \ll 1$ there exist supersymmetric vacua at relatively
small volume (the KKLT scenario \cite{kklt}). These represent different phases of
the landscape, and the location and properties of these phases depend on the values of the fluxes.

In this note we perform a systematic scan on the landscape for flux compactifications on the Calabi-Yau $\mbb{P}^4_{[1,1,1,6,9]}$.
 This Calabi-Yau manifold
provides one of the simplest concrete non-trivial examples of moduli
stabilisation. It
has
been very much studied over the past few years \cite{Candelas:1994hw,
  Denef:2004dm, hepth0502058} and has
become a prototypical example for explicit calculations both for
cosmological \cite{racetrack2},  phenomenological
\cite{hepth0505076, hepph0511162, aqs,hepth0605141, hepth0606047,
 Conlon:2006wz, 07040737, 07070105} and astrophysical
 issues \cite{cq}. It is then
interesting to fully explore the vacuum structure of this Calabi-Yau and
contrast with the partial results  found previously. In particular, the vacuum
structure can be only fully explored numerically. We perform such an
analysis here.
We scan across values of $W_0$ ranging over approximately twenty orders of magnitude, studying the different phases and properties of
the vacua that appear during this scan.
Several features emerge, for example the existence at small volumes of
a new class of vacua different from the standard KKLT and large volume minima.
We expect that the rich vacuum structure
of this manifold will also be shared in more complicated compactifications.

The organisation of this note is as follows. In section \ref{review} we give a brief review of the effects that enter the moduli potential
and the computation of soft supersymmetry breaking terms. In section \ref{ModelUsed} we describe the model we use ($\mbb{P}^4_{[1,1,1,6,9]}$) and
the analytic results on vacuum structure that are available. In section \ref{minstruc} we describe the scan over $W_0$ and the structure of the
different minima that are present. In section \ref{softtermsec} we study how the soft terms vary as we scan over $W_0$,
and in section 6 we conclude.

\section{Moduli Fixing and Soft Terms}
\label{review}

 We work in the effective supergravity limit of type $IIB$
string theory \cite{eff-sugra}. Its massless bosonic fields in ten
dimensions consist of the metric $g_{MN}$; two $2$-forms $B_{MN}$,
$C_{MN}$ with field strengths $H_3$ and $F_3$ respectively; a complex
dilaton/axion scalar field \be S=e^{-\phi} +i a \ee and a $4$-form
$C_4$ with self-dual field strength. Upon compactification the
$10D$ background metric
splits into a direct product of an $N = 1$ supersymmetric  $4D$ Minkowski spacetime and a $6D$ Calabi-Yau
orientifold. The Calabi-Yau  contains $2$-cycles,
their dual $4$-cycles, and $3$-cycles. The sizes of the $4$-cycles 
define the K\"ahler moduli fields
\be T_i = \tau_i + i \theta_i, \ee while those of the $3$-cycles
define the complex structure  moduli fields $U_a$. 

Fluxes thread the internal $3$-cycles $\Sigma_a$ and are quantised in
integral co-homology
\be \int_{\Sigma_a} H_3 = n_a, \quad \int_{\Sigma_b} F_3 = m_b, \quad
n_a, m_b \in \mathbb{Z}. \label{fquanta} \ee
They generate a K\"ahler moduli independent superpotential \cite{GVW}
\be W=\int G_3 \wedge \Omega, \ee where $G_3 = F_3 - iSH_3$ and
$\Omega$ is the holomorphic $(3,0)$-form of the internal space. This
fixes the dilaton and complex structure moduli fields
\cite{GKP}. After adding non-perturbative effects involving the
K\"ahler moduli, such as gaugino condensation, the
effective  superpotential takes the form \cite{kklt}
\be \hat{W} = W_0 + \sum_i A_i e^{-a_i T_i} \label{s-pot}.\ee
Here $ W_0 = \left< \int G_3 \wedge \Omega \right>$.
Together with these the four-dimensional effective theory is specified
by the K\"ahler potential at tree level
\be \hat{K} = -2\ln \mc{V} + K_{cs} \label{k-pot} \ee
where $\mc{V}$ is the Einstein-frame volume of the CY internal
space and $K_{cs}$
carries the complex structure moduli and the dilaton
dependence. It can be easily seen from here that the K\"ahler moduli
can be fixed at supersymmetric ($D_i W = 0$) minima of the $\mc{N}=1$
supergravity (scalar) potential
\be \label{n=1pot} V = e^{\hat{K}} \left[ \hat{K}^{i \bar{j}}
D_i\hat{W} D_{\bar{j}} \overline{\hat{W}} - 3 \vert \hat{W} \vert^2
\right]\ee where
\be D_i\hat{W}= \partial_i \hat{W} + (\partial_i \hat{K})\hat{W}
\textrm{ and } \hat{K}^{i \bar{j}} = (\hat{K}_{i \bar{j}})^{-1}.\ee

It is important to note that the fluxes (\ref{fquanta}) are
quantised in units of the
string scale $\alpha'$ and so naturally $|W_0|$ takes $\mc{O}(1)$
values. But since general Calabi-Yau manifolds have large numbers of
three cycles there are very many ways of turning on fluxes and by
tuning the flux integers (\ref{fquanta}) it is possible for $|W_0|$ to
take values arbitrarily close to zero \cite{hepth0404116}.\footnote{It
is however bounded above by an $\mc{O}(1)$ constant since the
magnitudes of fluxes themselves are bounded by tadpole cancellation
conditions.} The scanning performed in this note will vary $W_0$ as a smooth parameter. 
The K\"ahler potential also receives
$T_i$-dependent perturbative corrections that can be equally or more
important in stabilisation compared to the non-perturbative
corrections to $W$ \cite{hepth0502058, hepth0505076}. 

Supersymmetry breaking occurs if the moduli are fixed at non-supersymmetric
minima such as the large-volume minimum. Supersymmetry breaking is quantified
by the F-terms
\be \label {fterm} F^m = e^{\hat{K}/2} \hat{K}^{m \bar{n}} D_{\bar{n}}
\overline{\hat{W}}, \quad \hat{K}^{m \bar{n}} = (\partial_m
\partial_{\bar{n}} \hat{K})^{-1}. \ee
Given the F-terms for each class of minima 
the soft supersymmetry breaking terms are computed from the effective
supergravity Lagrangian density first by expanding the potentials in
powers of matter fields $C^\alpha$
\bea
W &=& \hat{W} + \mu H_1 H_2 + \frac{1}{6} Y_{\alpha
  \beta \gamma} C^{\alpha} C^{\beta} C^{\gamma} + \ldots, \\
K &=& \hat{K} + \tilde{K}_{\alpha \bar{\beta}}C^{\alpha}
C^{\bar{\beta}} + (Z H_1 H_2 + h.c.) + \ldots.
\eea
Here $H_{1,2}$ represent vector-like matter (in particular MSSM higgs bosons)
 and
$\tilde{K}_{\bar{\alpha} \beta}$ is the K\"ahler metric for matter
fields. Assuming a diagonal matter metric the
Lagrangian density can be written as
\be \mc{L}_{soft} = \tilde{K}_{\alpha} \partial_\mu C^{\alpha}
\partial^\mu \bar{C}^{\bar{\alpha}} - m_\alpha^2 C^{\alpha}
\bar{C}^{\bar{\alpha}} - \left[ \frac{1}{6} A_{\alpha \beta \gamma}
  \hat{Y}_{\alpha \beta \gamma} C^\alpha C^\beta C^\gamma +
  \ldots \right]\ee
with the scalar masses and A-terms given by \cite{bim}
\bea
m_\alpha^2  &=& m_{3/2}^2 + V_0 - F^{\bar{m}}F^n \partial_{\bar{m}}
\partial_n \log \tilde{K}_\alpha. \label{scalars} \\
A_{\alpha \beta \gamma} &=& F^m \left[\partial_m\hat{K} + \partial_m \log
  Y_{\alpha \beta \gamma} - \partial_m \log (\tilde{K}_\alpha
  \tilde{K}_\beta \tilde{K}_\gamma) \right] \label{atm}.
  \eea
The gravitino mass is given by
\be m_{3/2} = e^{\hat{K}/2}|\hat{W}|.\ee
The canonically normalised gaugino masses are
\be M_a = \half \frac{F^m \partial_m f_a}{ \hbox{Re} f_a},
\label{gmass} \ee
where $f_a$ is the gauge kinetic function whose form depends on whether the
gauge fields come from D3 or D7-branes: at tree level $f_{D3} = S,
f_{D7} = T$. In the following section \ref{ModelUsed} we describe the
Calabi-Yau manifold representing the internal space we use as an
explicit model for the soft terms calculations.

\section{The $\mathbb{P}^4_{[1,1,1,6,9]}$ Model}
\label{ModelUsed}
\paragraph{}
For concreteness we work with the well studied $\mathbb{P}^4_{[1,1,1,6,9]}$ Calabi-Yau
compactification with
 two K\"ahler moduli fields: $T_b$ and $T_s$ with the first
controlling the overall volume of the internal space and the second
one corresponding to the size of a smaller 4-cycle that may be
considered as a blow-up cycle. With the dilaton and complex structure
moduli fixed by fluxes the effective theory for the K\"ahler moduli is
\bea \label{pots}
 \hat{W} &=& W_0 + Ae^{-a T_s} + Be^{-b T_b},\\
\label{potsK}  \hat{K} &=& -2 \log{\left({\cal{V}} + \frac{\xi}{2}\right)}, \\
\textrm{where } {\cal{V}} &=& \frac{1}{9\sqrt{2}}(\tau_b^\frac{3}{2} -
\tau_s^\frac{3}{2}). \label{volm}\eea
Here $T_{b,s}= \tau_{b,s}+i\theta_{b,s}$ are the K\"ahler
moduli, $a,b = \frac{2 \pi}{N_{a,b}}$ for gaugino condensation on a
gauge group of rank $N_{a,b}$ and %$.\xi \sim \frac{1}{g_s^{3/2}}$
$\xi$ parametrises the $\alpha'^3$ correction to the K\"ahler moduli
space. The scalar potential takes the form
\bea \label{scalarV}
 V &=& e^{\hat{K}} \sum_{i,j} K^{i \bar{\jmath}}
 a_iA_i a_j \overline{A_j} e^{-(a_i T_i +
 a_j T_j)} \nonumber \\
 & & +
 (e^{\hat{K}} \sum_{i, j} K^{i \bar{\jmath}} (-a_i A_i)
 e^{- a_i T_i} \overline{\hat{W} (\partial_j \hat{K})} +
 c.c.) \nonumber  \\
 & & + 3\xi e^{\hat{K}} |\hat{W}|^2 \frac{\xi^2 + 7\xi {\cal{V}} +
 {\cal{V}}^2}{({\cal{V}} - \xi)(2{\cal{V}} + \xi)^2}.
\eea
This potential is singular at $\cal{V}= \xi$ and therefore constraints
the valid range of values of the volume to $\cal{V}> \xi$. 
We will scan over $W_0$ keeping all other parameters (including
$g_s$) constant. In the full theory different choices of fluxes would
also change the values of $A_i$ and $g_s$. The purpose of our
restriction to scanning simply over $W_0$ is that it makes it easy to
see the phase structure of minima without attempting to vary all
scales at the same time.

\subsection{Analytic Results}
%\subsection{$\mathbb{P}^4_{[1,1,1,6,9]}$ Vacua Classification}
\label{nosusy}
\paragraph{}
Depending on the values of flux the superpotential $W_0$ and the magnitude
of the gaugino condensation parameters $a$ and $b,$  different
types of vacua are realised from the $\mathbb{P}^4_{[1,1,1,6,9]}$
model. Here we present the general results about the vacuum structure
that can be derived analytically and leave the numerical results for
the following section.

The supersymmetric minima are easiest to study analytically since they are found by
solving the first order equations $D W = 0$ for each of the moduli
fields. For the two moduli case, we have \be D_s \hat{W} = D_b
\hat{W}= 0.\ee This implies:
\bea \label{susy}
a A e^{-a T_s} - \frac{3\hat{W}}{2 \mc V} \tau_s^{\half} &=& 0,
\nonumber \\
b  B e^{-b T_b} + \frac{3\hat{W}}{2 \mc V} \tau_b^{\half} &=& 0.
\eea
These give an AdS solution with the constraint that
\be W_0 = e^{-a \tau_s} \left[1 + \frac{2}{3} a \tau_s \right] -e^{-b
\tau_b} -\frac{a \xi }{3 \tau_s^{1/2}} e^{-a \tau_s} \ee at the
minimum of the potential. The depth of the potential at the minimum is
given by
\be V_0\ =\ -\frac{4}{3} \frac{a^2 e^{-2a \tau_s}}{\tau_s} =
-\frac{4}{3} \frac{b^2 e^{-2b \tau_b}}{\tau_b}. \label{klvac}\ee
Taking the ratio of equations in (\ref{susy}) gives
\be \label{susy2}
e^{(bT_b-aT_s)} =
-\frac{bB}{aA}\left(\frac{\tau_s}{\tau_b}\right)^{1/2}.\ee
By absorbing the phases of $A$ and $B$ into $T_s$ and $T_b$ it can be
seen that the angular phases $\theta_s, \theta_b$ are constrained to
satisfy $e^{ia\theta_s}=-e^{ib\theta_b}$. Also from equations
(\ref{susy}) it can be seen that the phase $e^{ia\theta_s}$ is the
same as the phase of the flux superpotential $W_0$ which without loss
of generality can be fixed to $-1$ so $\theta_s=\pi,$ $\theta_b=0$. We
are then left with the real parameters $\tau_s, \tau_b$ for which
we have two real equations. It is straightforward to extract the
following general conclusions from them.
\begin{enumerate}
\item
{\it There are no supersymmetric solutions with $a \ll b$:}
This is easily seen by rewriting equation (\ref{susy2}) into the form
\be b\tau_b-a\tau_s\ =\
\log\left[\frac{bB}{aA}\left(\frac{\tau_s}{\tau_b}\right)^{1/2}\right].
\ee This implies that $b\tau_b\sim a\tau_s$ at the minimum. But
$\tau_b > \tau_s$ is required in order to get positive volume and
therefore as long as $A$ and $B$ are comparable then $a > b$ must be
satisfied. In the case of a large hierarchy between $A$ and $B$
supersymmetric solutions may exist with $a$ slightly less than $b$.
\item
{\it There are no non trivial supersymmetric solutions for $W_0=0$:}
Using equations (\ref{susy}) and (\ref{susy2}) above it can be see
that for $W_0=0$ and $a>b$
\be
ab-\frac{3}{2\mc V}\left[b\tau_s^{1/2} - a\tau_b^{1/2}\right] = 0\ee
and substituting the expression for the volume from equation \ref{volm}
gives \be ab\left(\tau_b^{3/2}-\tau_s^{3/2}\right) +
3\left(a\tau_b^{1/2}- b\tau_s^{1/2}\right)\ =\ 0.\ee
This equation can never be satisfied because both terms are
positive. The trivial case of getting a supersymmetric solution with
$W_0=0$ is only possible at the decompactification limit.
\end{enumerate}
\paragraph{}
It follows that the supersymmetric KKLT class of vacua exist only for
$a>b$ and $W_0\neq 0$. In the one modulus case it was clear that
$W_0=0$ provides no solutions since a single exponential
superpotential gives a runaway behaviour, but in the many K\"ahler
moduli case this is not automatic because the sum of exponentials in
$W$ may have been in principle enough to generate a nontrivial
vacuum. Furthermore, as for the one modulus case, the supersymmetric
KKLT solutions can exist only as long as $W_0$ is small enough that
equations (\ref{susy}) may be solved with large enough $\tau_s,
\tau_b$ to keep the supergravity approximations valid. As from
equations (\ref{susy}) it follows that $\tau_b \sim \ln W_0, \tau_s
\sim \ln W_0$, as $W_0 \to 0$ the behaviour of the volume and
potential with $W_0$ is \be \mc{V} \sim ( - \ln W_0)^{3/2}, \qquad V_0
= -3 W_0^2. \ee
\FIGURE{\epsfig{file=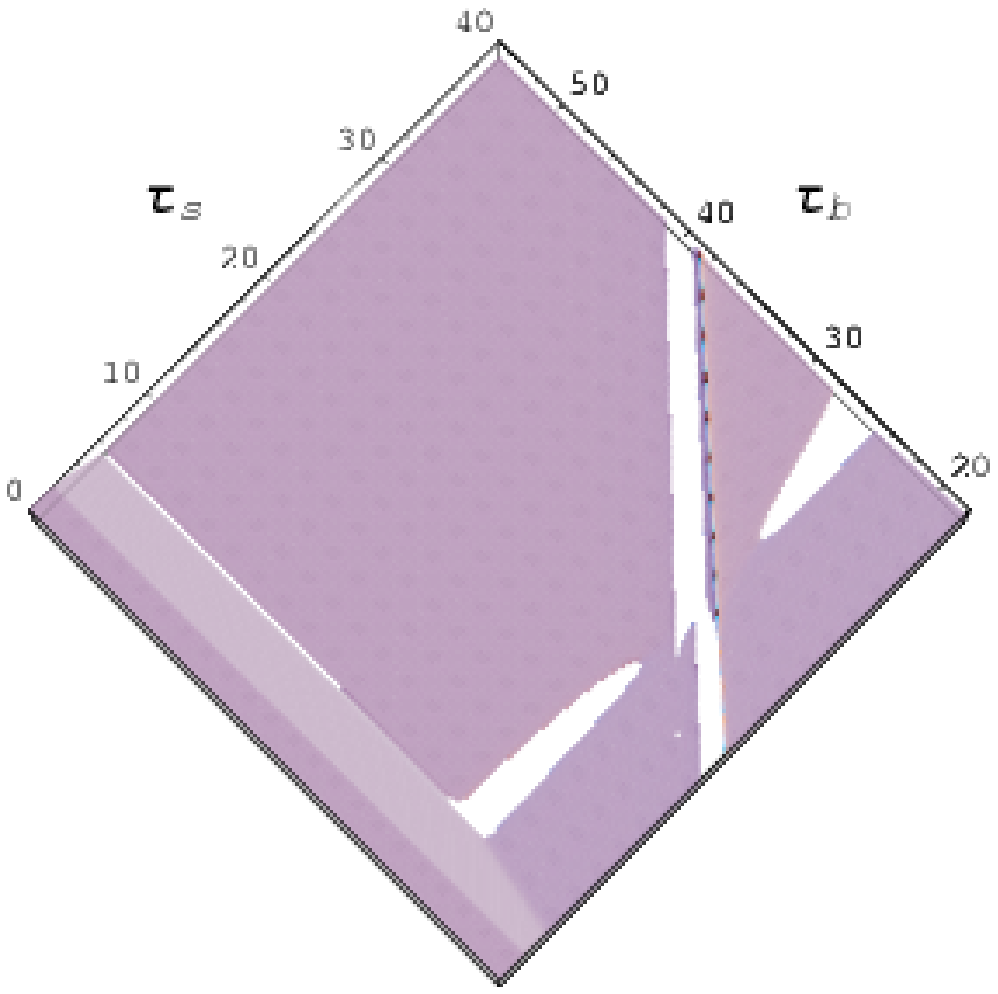,width=.6\textwidth}
        \caption[p]{\footnotesize{\it A $3D$ plot of $\log$
        of the potential (\ref{scalarV}) versus the two moduli fields
        $\tau_{b,s}$. On either side of the $\tau_b = \tau_s$
        white patch there are patches representing regions where the
        potential energy is negative and hence a valley where a
        minimum may be found. Any minimum in the valley
        to the right of $\tau_b = \tau_s$ is unphysical since it has
        $\mc V < 0$. The $\mc M_{LV}$ minima set is found in the
        vertical part of the almost L-shaped valley. The other minima
        sets, $\mc M_{KL}$ and $\mc M_{new}$, are found in
        the horizontal part of the L-valley.}} \label{valleys}}
%\paragraph{}
Besides the supersymmetric minimum $M_{KL}$, the scalar potential has another class of minimum $\mc
M_{LV}$ which has been found analytically. It corresponds
to non-supersymmetric AdS  and occurs at
exponentially large volume with $\theta_s = 0$ and $\tau_b \gg
\tau_s$. In the large volume approximation the potential has the form
\be \label{scalarp}
V = \sum_{i,j} C_{1} { \sqrt{\tau_s} e^{- 2a \tau_s}\over{\mc V}} -
\sum_i C_{2} {\tau_s e^{-a \tau_s} \over {\mc V}^2} + {C_3\over {\mc
V}^3}.\ee
The third term in (\ref{scalarp}) results from the $\alpha'$
correction to the K\" ahler potential equation(\ref{k-pot}) and the
$C$'s are constants of order unity. It can be
seen \cite{hepth0502058, hepth0505076} that $V$ has a minimum such
that one of the
moduli, $\tau_b$, is exponentially large and the other is $\tau_s
\gsim {\mc O}(1)$. The volume at the minimum is given by
\be {\mc V} \sim W_0\,  e^{a \tau_s}.\ee
For generic values of the flux superpotential $W_0\sim 1$ this volume
is exponentially large.  
\paragraph{}

In the sections that follow we will be more explicit concerning the
concrete range of parameters in which the supersymmetric AdS and the
non supersymmetric AdS minima exist. We will also demonstrate
numerically the existence of a new class $M_{new}$ of non-supersymmetric AdS and dS
vacua
in particular ranges of $W_0$ values. The de Sitter vacua obtained are analogous to those obtained in
\cite{hepth0408054} (see also \cite{hepth0611332, eva-ds}). 

The
plot in figure \ref{valleys} gives an instance of all three
minima coexisting.
\FIGURE{\epsfig{file=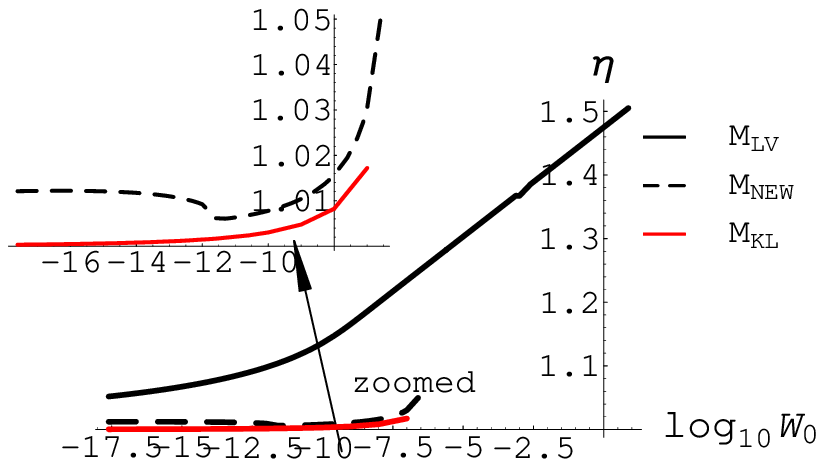,width=.6\textwidth}
        \caption[p]{\footnotesize{\it Check for supersymmetry breaking
        or preservation in the three classes of vacua. Supersymmetry
        is preserved if the ordinate $\eta=|V_0/3m_{3/2}^2|$ is
        exactly equal to $1$. The solid black line represents the
        ${\cal{M}}_{LV}$ set in which clearly supersymmetry is broken
        for all $W_0$. The red line is the KKLT minima set $\mc
        M_{KL}$ and its non-supersymmetric extension. The dashed 
        lines correspond to the ${\cal M}_{new}$ set of minima. Supersymmetry is
        always broken here also but at a smaller scale than
        for $\cal{M}_{LV}$.}}
    \label{susybrk}}

%%\section{Structure of the Different Minima} \label{minstruc}
\section{Structure of the Vacua Sets} \label{minstruc}
\paragraph{}
Depending on the non-perturbative and flux contributions multiple 
minima can coexist. The minima are differentiated by 
the volume of the internal space $\mc V$, the magnitude of the
bare cosmological constant $V_0$, the value of the axionic phase
$\theta$ at which the minimum settles and whether supersymmetry
breaking occurs or not. In order to study the variation of these quantities
with different flux choices we scan over the minima of the scalar
potential eqn(\ref{scalarV}) over different values of $W_0$. In
addition to this continuous scanning there is also a discrete
parameter choice, that heavily affects the structure of the minima,
coming from the relative magnitudes of the parameters $a$ and $b$.
For the numerical estimates we have chosen:
$$A = B = 1,\quad  \xi = 1.31, \quad a = 2\pi/0.85$$
 and,
depending on whether the case $a = b$ or $a > b$ is considered, $b =
\{2\pi/0.85, 2\pi/3.85\}$.

\subsection{Dependance on non-Perturbative Effects: $a$ and $b$
  Parameters}
\paragraph{}
We address the two cases $a > b$ and $a < b$ separately. 

\subsubsection{Case $a \leq b$}
\label{aleqb}
\paragraph{}
As already shown in section \ref{nosusy} no supersymmetric minimum
exists for these values of the parameters $a \leq b$. Scanning across
$W_0$, the only type of minima of the scalar potential found
corresponds to the set $\mc M_{LV}$ containing the large-volume
minimum. In the $W_0 \sim \mc O(1)$ region where the volume is
exponentially large, the scaling of the volume goes as $ \mc{V}
\sim W_0\, e^{a\tau_s}$. As $W_0$ is decreased, the volume decreases
linearly. This continues until $W_0$ is sufficiently small to
compensate the large exponential, such that the volume approaches
unity. As $W_0$ is further decreased, the minimum does not disappear
but instead exists in a small-volume phase. As far as we have been
able to check, this minimum continues to exist down to arbitrarily
small values of $W_0$. The variation of the volume with $W_0$ in this
minima set compared with the behaviour in the other sets (described in
section \ref{agtrb}) is shown in figure \ref{grav1}. In the
large-volume region, the behaviour
$\mc{V} \sim W_0$ is clear. After the transition to the small-volume
region, the internal volume is seen to be essentially constant and insensitive
to further decreases in $W_0$. It is possible to understand
analytically why these smaller volume minima exist in the 
 $\mc M_{LV}$ set at very small values of $W_0$. The large-volume
construction ensures that the potential goes to zero from below at
infinite volumes. However, the $\alpha'^3$ term in the scalar
potential diverges positively at small volumes. As the potential goes
to positive infinity at small volumes, and approaches zero from below
at large volumes, a minimum must exist somewhere in the intermediate
regime. Thus decreasing $W_0$ brings the large-volume minimum to smaller
values but does not destroy its existence.
\DOUBLEFIGURE[t]
         {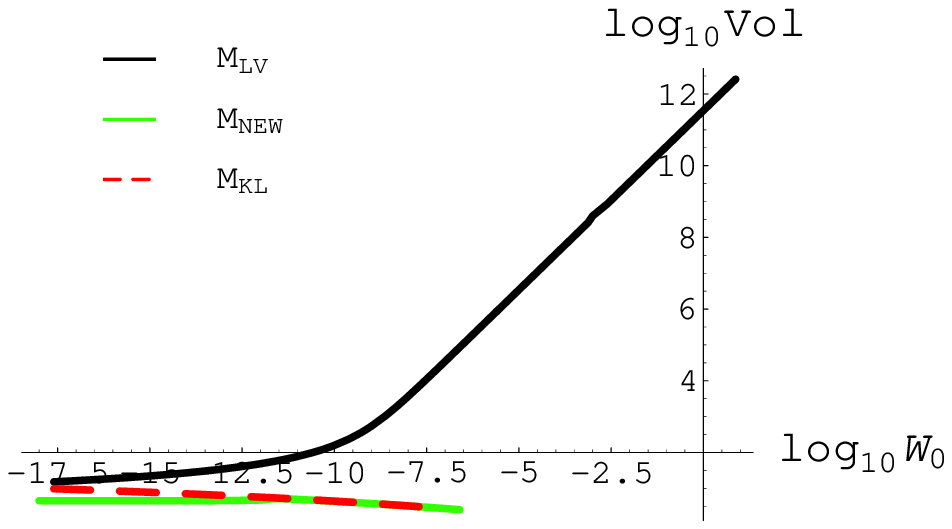, width=.4\textwidth}
         {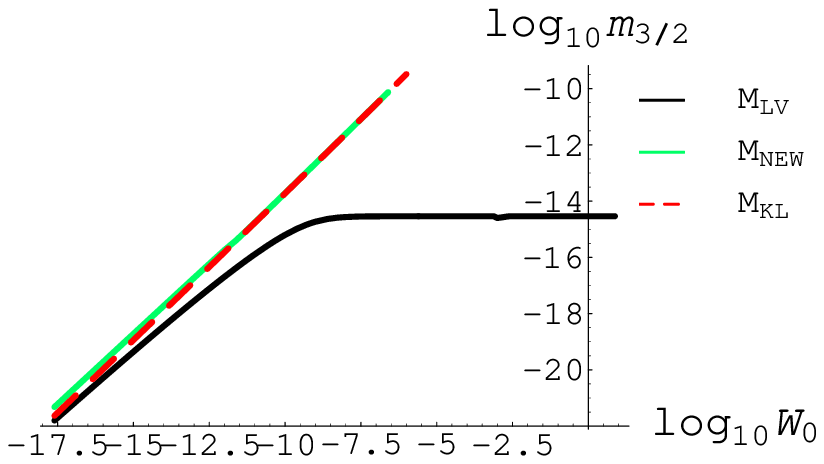, width=.4\textwidth}
         {\footnotesize{\it Plot shows the internal volume structure
         with various $W_0$ values in each of $\mc M_{LV}$ (black
         curve), $\mc M_{KL}$ (red line) and $\mc
         M_{new}$ vacua sets. The transition from a
         large to small volume phase is clearly seen at $\ln W_0
         \sim 10^{-9}$.}\label{grav1}}
         {\footnotesize{\it This plot shows the gravitino mass
         variation with $W_0$ for all the three vacua sets. In
         small $W_0$ regions corresponding to the small volume
         phase the mass is the same in all the minima
         types.}\label{grav}}

The gravitino mass $m_{3/2}$ determines the overall scale of the
supersymmetry breaking parameters. Figure \ref{grav} shows the
variation of the gravitino mass with $W_0$ in this minima set compared
with the behaviour in the other sets (described in section
\ref{agtrb}). The shape of the curve is as expected for the
large-volume minima: in the large volume
limit $m_{3/2} = \frac{W_0}{\mc{V}}$ and $\mc{V} \sim W_0$, making the
gravitino mass independent of $W_0$. Once $W_0$ is sufficiently
small, the volume of the minimum remains fixed and small. However
$W_0$ continues to decrease and thus the gravitino mass decreases
linearly with $W_0$.

The transition from large volume to small volume phases is best seen from the structure
of the cosmological constant with $W_0$ in the $\mc M_{LV}$  vacua
set compared with the behaviour in the other sets (described in section
\ref{agtrb}) as shown in figures \ref{vacc1} and  \ref{vacc}. The
figures clearly show the existence of a minimal value for the
cosmological constant around $W_0 \sim 10^{-9}$. This feature can be
analytically explained: For larger values of $W_0$, we are in the
large-volume region for which the minima can be found analytically with
an AdS cosmological constant
$$V_0 \sim \frac{W_0^2}{\mc{V}^3} \sim \frac{1}{W_0}.$$
As $W_0$ decreases, $\vert V_0 \vert$ therefore increases linearly, as
indeed seen in the figure \ref{vacc1}. Once in the small volume region,
$\mc{V}$ no longer has significant dependence on $W_0$, and so
$$V \sim \frac{W_0^2}{\mc{V}^3} \sim W_0^2$$ so the magnitude of $V_0$
decreases quadratically with $W_0$. This change is seen in
figure \ref{vacc} through the change in slope and direction. These minima are non-supersymmetric 
for all values of $W_0$. 
\FIGURE{\epsfig{file=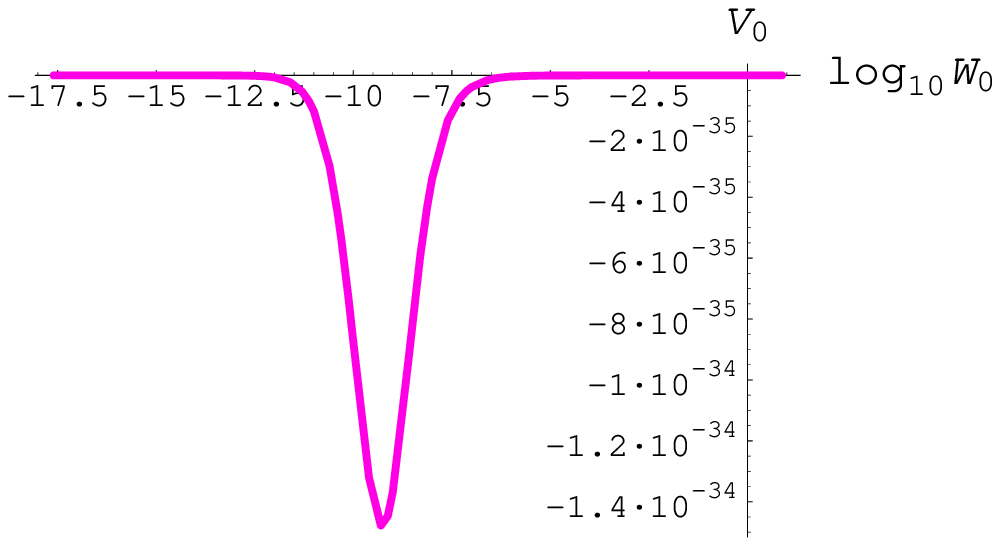, width=.6\textwidth}
        \caption[p]{\footnotesize{\it The variation of the bare
         cosmological constant $V_0$ with $W_0$ for the large-volume
         minimum. The minimum value of $V_0$ lies around $W_0 \sim
         10^{-9}$ and marks the phase transition.}} \label{vacc1}}

\FIGURE{\epsfig{file=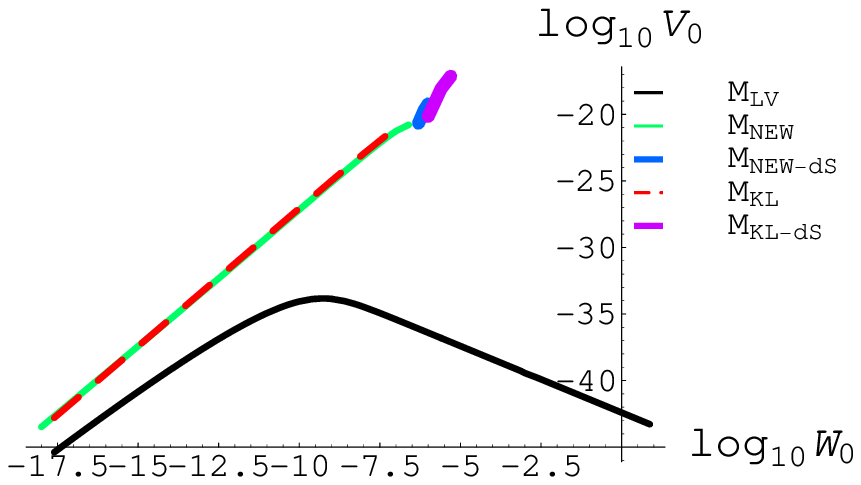,width=.6\textwidth}
        \caption[p]{\footnotesize{\it This plot shows the structure of
         $V_0$ over $W_0$ in the three vacua sets. The depth of the
         potential at the different minima or the cosmological
         constant is identical for the $\mc M_{KL}$ and  $\mc
         M_{new}$ vacua set. The minima in these sets become
         de Sitter in a small interval of $W_0$ values. The depth of
         the potential (now positive) in this region is shown by the
         bolder lines. The black curve is a {\it log - log} plot of
         $V_0$ for the $\mc M_{LV}$ vacua set. The slopes on either
         side of the turning point, $+2$ and $-1$, show the different
         structure of $V_0$ on $W_0$  from large and small
         volume regions.}} \label{vacc}}

\subsubsection{Case $a > b$} \label{agtrb}
\paragraph{}
In the $a > b$ scenario the potential eqn(\ref{scalarV}) has greater
richness of minima. Four distinct classes of vacua  can be identified 
for different values of $W_0$. We enumerate these and
describe their properties:
\begin{enumerate}
\item The large-volume minima $\mc M_{LV}$. 
These are realised for sufficiently large values of
  $W_0$ and have a similar structure to the case $a\leq b$ of the
previous section.
  $\tau_s$ is stabilised by effects non-perturbative in
  $\tau_s$ and $\tau_b$ is stabilised by effects perturbative in
  $\tau_b$. With the superpotential parameters $A, B$ real
  and positive and $W_0$ real and negative the minima settle on the
   at $\theta_s = 0$ with $\theta_b$ undetermined up to non-perturbative effects that play no role
  in stabilising $\tau_b$.
\item The supersymmetric (KKLT) minimum \cite{kklt}, which exists at
  small volume ($\mc{V} \sim \ln W_0$) for sufficiently small values
  of $W_0$. It has $$a \tau_s \sim \ln W_0, \qquad b \tau_b \sim \ln
  W_0.$$ Here $\tau_s$ is stabilised by effects non-perturbative in
  $\tau_s$ and $\tau_b$ is stabilised by effects non-perturbative in
  $\tau_b$. With the above choice of sign conventions, $\theta_s = \pi$ and
  $\theta_b = 0$.
\item A new set of minima $\mc M_{new}$ coexisting in the
  space of parameters with the supersymmetric $\mc M_{KL}$ with
  $\theta_b = 0$ and $\theta_s$ is undetermined. This set
  has some similar properties to $\mc M_{KL}$ but is always
  non-supersymmetric as shown in figure \ref{susybrk}. 
  The structures of gravitino mass, the internal space
  volume and the cosmological constant $V_0$ in these sets of minima
  are similar as can be seen in the corresponding
  plots of the mentioned quantities in figures \ref{grav1}, \ref{grav}
  and \ref{vacc}. The minima settle such that $$ b \tau_b \sim \ln W_0,
  \qquad \ln W_0 \ll a \tau_s, \qquad \tau_s < \tau_b.$$ $\tau_b$ is
  stabilised by effects non-perturbative in $\tau_b$, whereas $\tau_s$
  is stabilised by effects perturbative in $\tau_s$. The origin of
  this kind of minima can be understood as follows. Suppose we first
  neglect the existence of the $\alpha'^3$ corrections, and also take
  $\tau_s$ to be sufficiently large that the terms non-perturbative in
  $\tau_s$ can be ignored. In this case the only terms contributing to
  the scalar potential are those non-perturbative in $\tau_b$. In this
  limit the effective scalar potential is
\be V = \frac{\tau_b^2 a_b^2 A_b^2 e^{-2 a_b \tau_b}}{\mc{V}^2} -
  \frac{\tau_b a_b A_b W_0 e^{-a_b \tau_b}}{\mc{V}}. \ee
  With this $\tau_b$ is fixed as in KKLT to generate an effective
  negative potential, \be V \sim - \frac{W_0^2}{\mc{V}^2}.\ee
  The volume is a function both of $\tau_s$ and $\tau_b$ and is thus
  not stabilised (as $\tau_s$ was not stabilised). This potential can
  be decreased by decreasing the volume. To stabilise $\tau_s$ the
  $\alpha'$ correction has to be turned on. The appearance of the
  volume in the potential therefore generates a potential for
  $\tau_s$, which starts increasing (so as to decrease the volume and
  thus the potential). However, the $\alpha'^3$ correction diverges
  positively at small volume, giving a total potential \be V \sim -
  \frac{W_0^2}{\mc{V}^2} + \frac{\xi W_0^2}{\mc{V}^2 (\mc{V} -
  \xi)}\ee so at some point this potential must turn around at small
  volume giving a non-supersymmetric minima at small volume.
  As this minimum exists at very small volume (significantly smaller than in
KKLT), it is not clear that the supergravity arguments leading to its existence
can be trusted in the full string theory.

\item A de Sitter (dS) set of minima contained in the $\mc M_{KL}$ vacua
  set that are connected with other supersymmetric (KKLT) minima since
  one can move from the former to the latter set (and vice versa) by
  smoothly varying $W_0$.  These are the 2-modulus versions of the de Sitter minima
  of \cite{hepth0408054}.
  They exist for a very small range of $W_0$, where
  perturbative and non-perturbative corrections balance each other, in
  both $\mc M_{KL}$ and $\mc M_{new}$ vacua sets.
\end{enumerate}
\subsection{Dependence on Fluxes: $W_0$ Parameter}
\paragraph{}
Varying the flux superpotential $W_0$ four distinct ranges with
different vacua structure can be identified. These ranges are
illustrated schematically in figure \ref{w0line} and explained in the
remaining part of this subsection.

\FIGURE{\epsfig{file=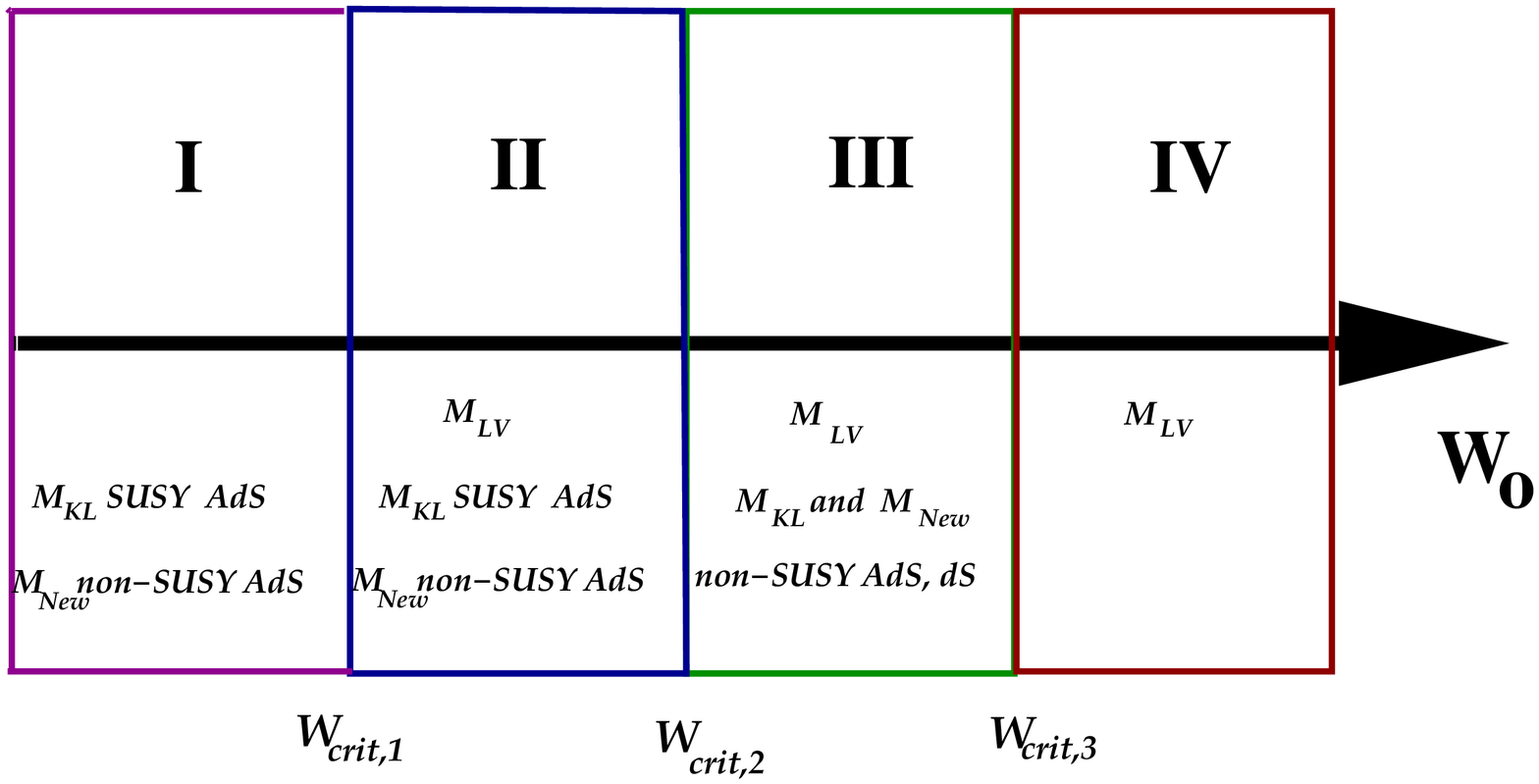,width=0.9\textwidth}
        \caption[p]{\footnotesize{\it Four $W_0$ regions with different
        vacua sets for $a>b$. The large-volume set $\mc
        M_{LV}$ is the most generic, it is always
        non-supersymmetric AdS and disappears after `merging'
        with the supersymmetric $\mc M_{KL}$ set at very small values
        of $W_0$.
 The KKLT set in $\mc M_{KL}$ and the new $\mc
        M_{new}$ set exist in the regions with very
        small $W_0$ and evolve into non-supersymmetric, AdS and
        then dS minima before disappearing altogether for larger
        values of $W_0$ as described in text (section \ref{agtrb}).}}
    \label{w0line}}
\subsubsection{Extremely small $W_0$, $W_0 \leq W_{crit,1}$}
\paragraph{}
For extremely small values of $W_0 \leq W_{crit,1}$ there exist only two
vacua sets of the potential eqn(\ref{scalarV}): $\mc{M}_{KL}$ and
$\mc{M}_{new}$. Both are AdS, with the KKLT set being
supersymmetric and the new set being non-supersymmetric. The potential
in these minima sets essentially settles at the same depth. We define
this range as region I.
\subsubsection{Small $W_0, W_{crit,1} \leq \vert W_0 \vert \leq
  W_{crit,2}$}
\paragraph{}
Starting with the two minima set of the previous region I and as $W_0$
is increased to $W_{crit,1} \sim 10^{-17}$ a third minima type
appears. This corresponds to the non-supersymmetric large-volume
minima set $\mc M_{LV}$. It is distinguished from the other ones by
the $\theta_s$ axion phase. Initially the minima
in this class appear at values of $\tau_s$ and $\tau_b$ very similar
to those in $\mc M_{KL}$. But as $W_0$ is further increased, the
volume rapidly increases and thus the minima separate leaving the
other ones in the small ($\tau_s$, $\tau_b$) region. 
\subsubsection{Small $W_0, W_{crit,2} \leq \vert W_0 \vert \leq
  W_{crit,3}$}
\paragraph{}
$W_{crit,2}$ is defined as the
maximal value of $W_0$ for which supersymmetric solutions 
are possible. As $W_0$ is increased beyond this point all
the minima are non-supersymmetric because the equation $D_i W = 0$ can
no longer be solved. 

For our model with two moduli fields
\be D_iW = 0 \rightarrow W_0 = -\left[ \sum_{j=s,b} A_j e^{-a_j T_j} +
  \frac{(\mc V + \xi/2)}{t^i} a_i A_i e^{-a_i T_i}
  \right] \label{susycheck} \ee
where we have used $\partial_{T_i}K = -t^i/(\mc V + \xi/2)$, where $t^i$
measures the area of 2-cycles with $$\tau_k = Re(T_k) = \partial_{t^k}
\mc V = \half \kappa_{ijk} t^i t^j, \qquad \mc V = \frac{1}{6}
\kappa_{ijk} t^i t^j t^k.$$ Note also that we require $\mc V > \xi >
0$ so $t^i \neq 0$. From eqn(\ref{susycheck}) $t^i \rightarrow 0$ will
imply $W_0 \rightarrow \infty$ and hence supersymmetry cannot be
preserved. For the symmetry to be preserved $t^i$ must be bounded away
from zero. Hence for any supersymmetric minimum the
possible $W_0$ values are bounded from above, $W_0 \leq W_{max}$,
since larger $W_0$ require smaller $t^i$. In this manner any minimum
with $W_0 > W_{max}$ will be non-supersymmetric. This explains why
supersymmetry is broken in $\mc M_{KL}$ for $W_0 \gtrsim W_{max} =
W_{crit,2}$.

As $W_0$ is further increased, the minima in the $\mc M_{KL}$ and $\mc
M_{new}$ vacua set become de Sitter before disappearing altogether at $W_0
\sim W_{crit,3}$. 
\FIGURE{\epsfig{file=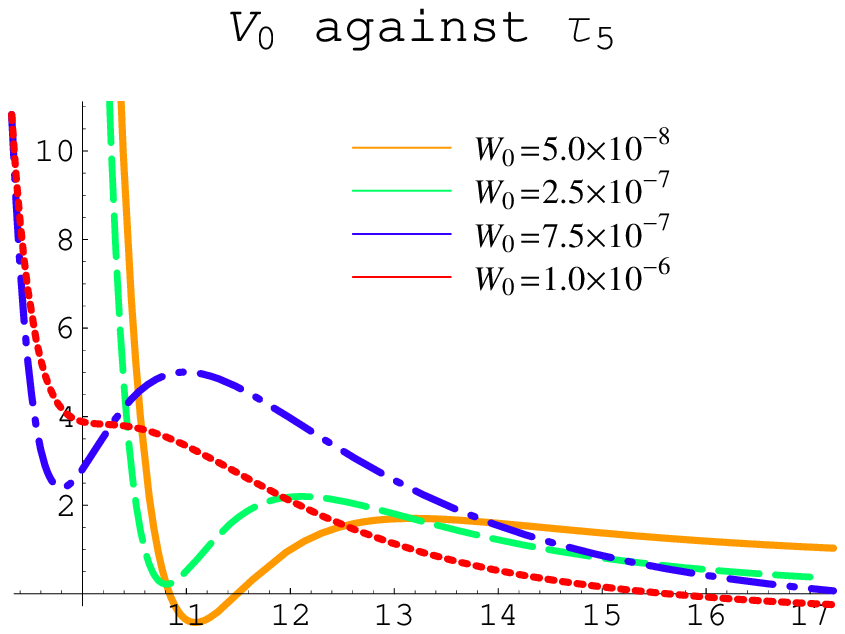,width=.6\textwidth}
        \caption[p]{\footnotesize{\it The behaviour of a minimum which
        becomes dS before finally disappearing altogether as $W_0$ is
        increased in the range $W_{crit,3} >W_0 > W_{crit,2}$.}}
    \label{ds1to4}}
    As shown in figure \ref{ds1to4} the minimum disappears
 when the positive term in $V$ becomes dominant. The range of $W_0$ over which the minima go
de Sitter without introducing uplifting effects is rather
small. 
\subsubsection{Large $W_0, W_{crit,3} \leq \vert W_0 \vert $}

For $W_0 \gtrsim W_{crit,3}= 2.5 \times 10^{-6}$ only minima in the
large-volume phase  $\mc M_{LV}$  survive. The
small-volume phases and all other vacua type no longer exist.

The soft term structure is described in the next section \ref{softtermsec}.

\section{Soft Terms Structure in the Vacua Sets}
\label{softtermsec}
\DOUBLEFIGURE[t]
         {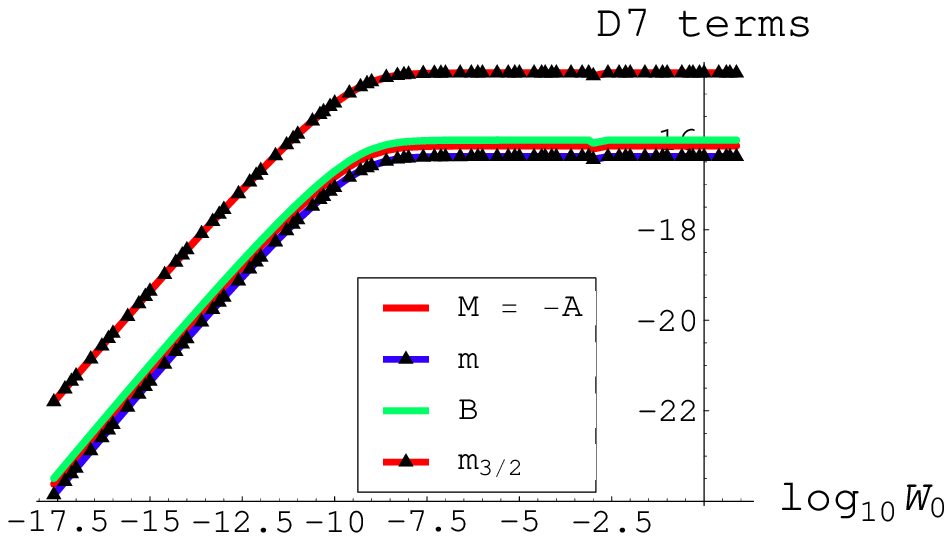, width=.4\textwidth}
         {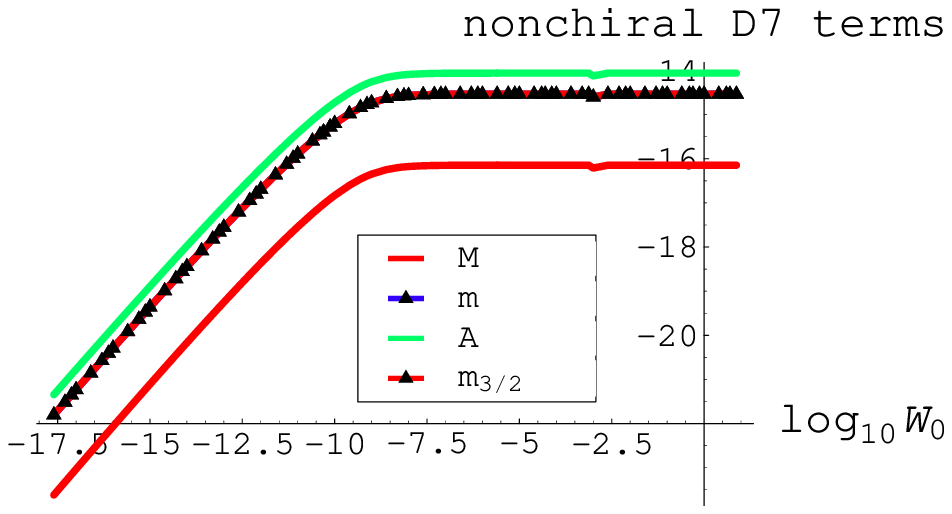, width=.4\textwidth}
         {\footnotesize{\it The dependence of soft terms for chiral
     D7-brane matter on $W_0$ for ${\cal{M}}_{LV}$. The terms are degenerate and
     suppressed with respect to the gravitino mass. The amount of
     supression is constant over all values of
     $W_0$.}\label{scape}}
         {\footnotesize{\it Structure of soft terms for non-chiral
         D7-brane matter for ${\cal{M}}_{LV}$.
 The gaugino mass is reduced with respect to,
     while the scalar mass is degenerate with the gravitino
     mass. The $A$-term is slightly greater than $m_{3/2}$. Unlike
     the D3 soft terms, the small hierarchy between the terms
     remains constant for all $W_0$.}\label{d7-2}}

\DOUBLEFIGURE[t]
         {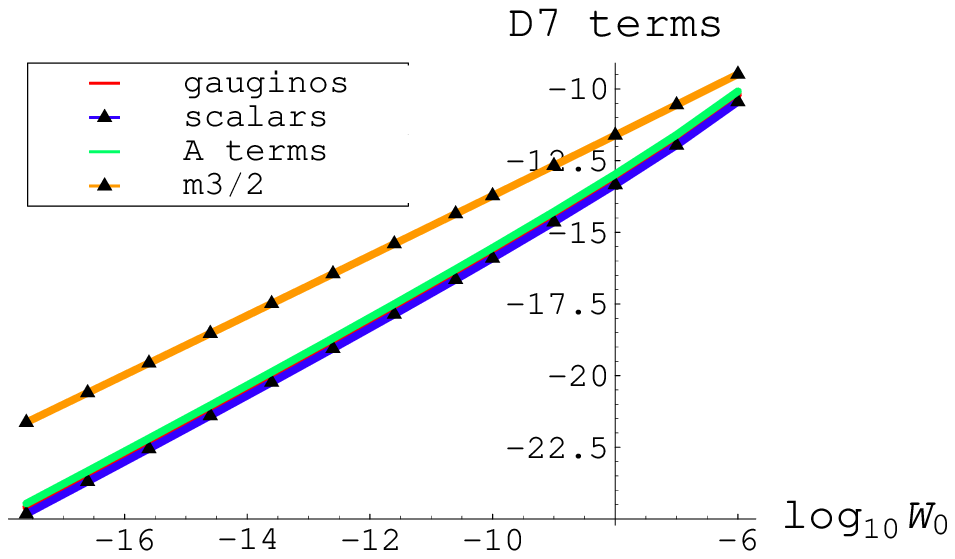, width=.4\textwidth}
         {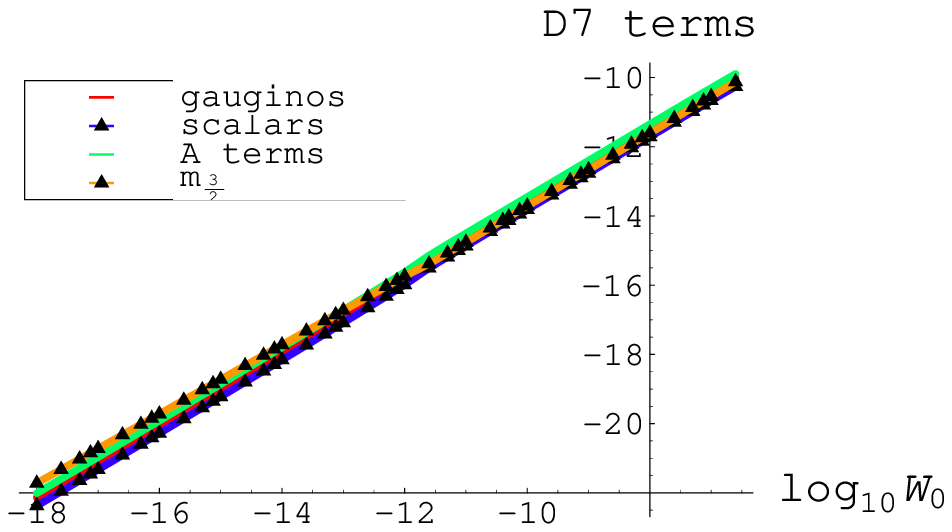, width=.4\textwidth}
         {\footnotesize{\it The dependence of soft terms for chiral
     D7-brane matter on $W_0$ for ${\cal{M}}_{KL}$. Here the 
behaviour is different from the ${\cal{M}}_{LV}$ case as the soft
         terms scale as the gravitino mass with  $W_0$.}\label{d7-3}}
         {\footnotesize{\it Structure of soft terms for non-chiral
         D7-brane matter for ${\cal{M}}_{new}$. In this case all the
         soft terms scale as the gravitino mass.}\label{d7-4}}

In this section we present the behaviour of soft terms as $W_0$ is
varied. As in the $a\le b$
case there is only one class of vacua, the $\mc M_{LV}$ set, which is also present in the
$a \ge b$ case. Therefore,  we only address  the latter ($a>b$) case. 
We focus on D7 soft terms as these are more phenomenologically relevant.
\subsection{D7 soft terms}
For matter fields in the bifundamental representation, arising from
open strings stretching between magnetised D7-branes, the soft terms
have been computed in \cite{Conlon:2006wz} for the large-volume model
case of $W_0 \sim \mc O(1)$ as $m:M:A \sim 1: \sqrt{3}: -1$ with $m
\sim m_{3/2}$. To obtain this result it was assumed that the D7 branes
carrying the MSSM are wrapping the small 4-cycle, which we will also
assume here. For the large volume phase, this is required to
obtain gauge couplings of the correct magnitude, since branes wrapping
the large 4-cycle would yield gauge couplings far too small to be
compatible with the observed values.
\paragraph{}
The numerically obtained soft terms structure in all of $\mc M_{LV}$,
$\mc M_{KL}$ and $\mc M_{new}$ vacua sets over the different
$W_0$ ranges is shown in figure \ref{scape}. There are basically two
different soft terms characteristics in $\mc M_{LV}$ over all the
scanned $W_0$ values. Above the turning point $W_0^c \sim 10^{-9}$ the
soft terms are independent of $W_0$. Below this critical value the soft
terms vary linearly with $W_0$. In both cases the mass hierarchy
between the soft terms $m, M, A$ and the gravitino is as
predicted in \cite{hepth0605141,Conlon:2006wz} for the large-volume
model. The kind of mass hierarchy obtained is different for the case
of D7 matter fields in the adjoint representation, as shown in figure
\ref{d7-2}. The main difference is that here only the gaugino mass
term is suppressed compared to the gravitino mass. This is due to the
fact that the F-term of the small K\" ahler modulus, $F^s$, is
suppressed with respect to the gravitino mass by a factor $\log
(M_P/m_{3/2})$ as described in \cite{hepth0605141}.
\paragraph{}
For both the ${\cal{M}}_{KL}$ and ${\cal{M}}_{new}$ sets the soft
terms for chiral matter
scale with $W_0$ very different from the ${\cal{M}}_{LV}$ models
as it can
be seen in figures \ref{d7-3} and \ref{d7-4}.
For ${\cal{M}}_{KL}$  the gravitino mass is hierarchicaly larger than
all the soft terms, but the difference increases slightly with
decreasing $W_0,$ whereas for ${\cal{M}}_{new}$ all the soft terms are
degenerate with the gravitino mass.

%\subsection{D3 soft terms:}
%For matter on D3 branes, the soft terms were obtained in \cite{hepth0505076}
%as  $m:M:A \sim 1/{\mc V}^{1/6}: 1/\mc V: 1/{\mc V}^{4/3}$.
%In the large-volume case of $W_0 \sim \mc O(1)$, this
%is similar to a split supersymmetry scenario.
%The scalar masses and A-terms are independent of
%$W_0$ in the region where $\mc V \propto W_0$. On the other hand, the
%gaugino mass term is dependent on $W_0$ in all regions. The above
%ratio remains the same but with the amount of split between the scalar
%and gaugino masses decreasing since $\mc V$ decreases as $W_0$ is
%reduced. Below the critical value $W_0^c \sim 10^{-9}$,
%the volume $\mc V$, and hence
%the little hierarchy that remains, remains constant with $W_0$, as
%shown in figure \ref{d3termsfig}.
%\FIGURE{\epsfig{file=D3tms.eps,width=.4\textwidth}
%        \caption[p]{\footnotesize{\it Structure of soft terms for
%         $\mc M_{LV}$ minima and D3-brane matter fields.
%     The scalar mass is
%         degenerate with the gravitino mass while the gaugino
%         mass is suppressed with respect to the
%         gravitino mass by a power of the volume.}}
%             \label{d3termsfig}}
\section{Conclusions}

In this paper we have performed, for a particular Calabi-Yau,
 a scan over the different phases of the landscape by varying $W_0$. We have varied
 $W_0$ by twenty orders of magnitude, from $\mc{O}(1)$ to $\mc{O}(10^{-20})$, and have studied the
 structure and properties of the various minima.

A total of four different classes of minima were observed in the scan. These minima
can be either supersymmetric or non-supersymmetric, dS or AdS, and at large and small volume.
The supersymmetry breaking scales can be either high or low.
 We have plotted the variation of physical quantities such as the gravitino mass or the cosmological
 constant with $W_0$, and have been able to give analytic explanations for its behaviour.
 We have seen that certain minima can only exist in a restricted range of parameter space. For example,
the large volume solution can cease to be present at sufficiently
small values of $W_0$, turning into the KKLT solution,
whereas the non-supersymmetric de Sitter minima analogous to those
found in \cite{hepth0408054} are only present for a small range of $W_0$.

Some implications and potential applications of our results may be considered
\begin{enumerate}
\item{}
Our mostly
 numerical analysis complements the analytical results that have been
obtained before for this Calabi-Yau.
 It reproduces them in the regimes where the
analytical results are valid but it uncovers previously unnoticed
facts in the regimes where only the numerical analysis is available.
Examples are the existence of the new class of non-supersymmetric AdS and
dS minima and the soft terms behaviour for the large volume class in the
domain of small flux superpotential. We expect a similar structure of
minima to be present in more general Calabi-Yau compactifications.

\item{} The minima which are more reliable in this effective field theory
  analysis are the ones corresponding to very
large volume. It is reassuring that in a large
  region of the landscape the soft supersymmetry breaking parameters
  remain constant and then their value can be considered generic
  within the landscape.

\item{}
The KKLT and new minima exist for very small values of $W_0$
(and only for $a>b$) and are therefore less generic. They are also
less reliable as long as the volume is relatively small, especially in
the  non-supersymmetric cases.
 Nevertheless, their existence within the effective field
theory leads naturally to potential implications. Being
non-supersymmetric, the new minima offer a pattern of soft supersymmetry breaking
terms much different form the well studied large volume case
(and the KKLT case once lifting is included). The change from
supersymmetric AdS to non-supersymmetric AdS to dS was already noted
for the one modulus case in \cite{hepth0408054}. Here we have seen that the
new minima follow a similar behaviour. It may be interesting to
explore the cosmological implications and vacuum transitions
of this structure especially in regimes where the three classes of minima coexist once a lifting is
also included.

\item{}
For the large volume minima in the regime where $W_0$ is small enough
to cancel the large exponential dependence of the volume, the
phenomenological implications are very different from the
actual  large volume regime that has been studied in the literature.
As figure 3 illustrates, the gravitino mass stops being constant at a
critical  $W_0=W_c$ and then decreases with decreasing $W_0$. This
allows to consider phenomenological scenarios where the string scale
(independent of $W_0$) is large but the gravitino mass is a TeV
(similar to the KKLT case studied in \cite{choi}). It may be
interesting to explore the low-energy structure of soft terms for
`benchmark points' corresponding to the domain where $W \ll W_c$ and
also at $W\sim W_c$ which will have distinctive physical behaviour
than the $W> W_c$ studied already in detail all the way to LHC
energies
(see for instance
\cite{Conlon:2006wz, cksaq}).

\end{enumerate}

Our results lend support to the notion that
the string landscape has a rich structure of vacua with different
physical properties, and makes more pressing the task of determining which vacua have the
necessary properties to be phenomenologically successful.

\acknowledgments{} We thank B. Allanach, C. Burgess, M. Cicoli,
D. Cremades, F. Denef, F. Feroz, M. P. Garcia del Moral,
D. Grellscheid, S. Kom and D. Mackay for discussions. SA is supported
by Gates Cambridge Scholarship Trust and a member of St Edmunds
College, Cambridge. JC and KS are funded by Trinity College,
Cambridge. FQ is partially funded by SFC and a Royal Society Wolfson
Merit Award.

\end{document}